%% file: report.tex
\documentclass[11pt]{article}

\usepackage{alltt}

\newcommand{\back}{$\mathsf{\backslash}$}

\newcommand{\lb}{$\lbrace$}
\newcommand{\rb}{$\rbrace$}
\newcommand{\hgt}{$\rangle$}
\newcommand{\hlt}{$\langle$}

\newcommand{\code}[1]{{\sf{#1}}}
\newenvironment{program}{%
   \begin{alltt}}{%
   \normalsize\end{alltt}}

\newcommand{\Com}[1]{{\sl{#1}}}
\newcommand{\com}[1]{{\sl{#1}}}
\newcommand{\jml}[1]{#1}
\newcommand{\mr}[1]{\textsf{\em{#1}}}
\newcommand{\mc}[1]{\textsf{\sf\em{#1}}}
\newcommand{\id}[1]{\textsf{\sf{#1}}}
\newcommand{\persp}[1]{\underline{\smash{#1}}}

\begin{document}

\thispagestyle{empty}

\hspace*{-3em}
\begin{minipage}{6cm}
{\bf\sf\Huge IMAG}

\vspace*{2ex}

\begin{tabular}{@{}l}
{\bf\sf Institut d'Informatique et de} \\
{\bf\sf Math\'ematiques Appliqu\'ees} \\
{\bf\sf de Grenoble}
\end{tabular}
\end{minipage}

\vfill

\begin{center}
{\bf\sf\Huge LSR}
                                                                                
\vspace*{2ex}
                                                                                
{\bf\sf\Large Laboratoire Logiciels, Syst\`emes, R\'eseaux}
\end{center}
                                                                                
\vfill
                                                                                
\begin{center}
\begin{minipage}{9cm}
\begin{center}
{\bf RAPPORT DE RECHERCHE}
                                                                                
\vspace*{1.5ex}
{\bf Jartege: a Tool for Random Generation of Unit Tests
for Java Classes}
                                                                                
\vspace*{1.5ex}
{\em Catherine Oriat}
\end{center}
{\bf RR 1069} \hfill {\bf Juin 2004}
\end{minipage}
\end{center}

\vspace*{1ex}
                                                                                
\begin{center}
{\sf B.P. 72 -- 38402 SAINT MARTIN D'HERES CEDEX -- France}

\vspace*{1ex}

{\sf\small
Centre National de la Recherche Scientifique 

Institut National Polytechnique de Grenoble 
                                                                                
Universit\'e Joseph Fourier Grenoble I}
\end{center}

\newpage
\thispagestyle{empty}

\begin{center}
\bf\Large 
   Jartege: a Tool for Random Generation of Unit Tests
   for Java Classes
\end{center}

\begin{center}
Catherine Oriat

LSR-IMAG, Grenoble 

email: {\tt Catherine.Oriat@imag.fr}
\end{center}

\subsubsection*{R\'esum\'e}
   Ce rapport pr\'esente Jartege, un outil qui permet la g\'en\'eration
   al\'eatoire de tests unitaires pour des classes Java sp\'ecifi\'ees
   en JML. JML (Java Modeling Language) est un langage de sp\'ecification
   pour Java qui permet d'\'ecrire des invariants pour des classes, 
   ainsi que des pr\'e- et des post-conditions pour des op\'erations.
   Comme dans l'outil JML-JUnit, nous utilisons les sp\'ecifications
   JML d'une part pour \'eliminer des cas de test non pertinents,
   et d'autre part comme oracle de test. Jartege g\'en\`ere 
   de fa\c con al\'eatoire des cas de test, qui consistent
   en une s\'equence d'appels de constructeurs et de m\'ethodes 
   des classes sous test. L'aspect al\'eatoire de l'outil
   peut \^etre param\'etr\'e en associant des poids aux 
   classes et aux op\'erations, et en contr\^olant le 
   nombre d'instances cr\'e\'ees pour chaque classe sous test.
   L'utilisation pratique de Jartege est illustr\'ee par 
   une petite \'etude de cas.

\subsubsection*{Mots-cl\'es}
   Test, test unitaire, g\'en\'eration al\'eatoire de cas de test,
   Java, JML

\subsubsection*{Abstract}
   This report presents Jartege, a tool which allows random
   generation of unit tests for Java classes specified in JML.
   JML (Java Modeling Language) is a specification language
   for Java which allows one to write invariants for classes,
   and pre- and postconditions for operations.
   As in the JML-JUnit tool, we use JML specifications on
   the one hand to eliminate irrelevant test cases, and on
   the other hand as a test oracle.
   Jartege randomly generates test cases, which
   consist of a sequence of constructor and method calls for
   the classes under test.
   The random aspect of the tool can be parameterized by
   associating weights to classes and operations, and by
   controlling the number of instances which are created
   for each class under test.
   The practical use of Jartege is illustrated by a small
   case study.

\subsubsection*{Keywords}
   Testing, unit testing, random generation of test cases, Java, JML

\newpage

\cleardoublepage

\begin{center}
\bf\Large
                                                                                
Jartege: a Tool for Random Generation

of Unit Tests for Java Classes
\end{center}

\begin{center}
Catherine Oriat
                                                                                
LSR-IMAG, Grenoble
                                                                                
email: {\tt Catherine.Oriat@imag.fr}
\end{center}

\section{Introduction}

Main validation technique in software engineering, program testing
aims at ensuring that the program is {\em correct},
i.e. conforms to its specifications.
As the input domain of a program is usually very large or
infinite, exhaustive testing, which consists in testing the program
for all its possible inputs, is in general impossible.
The objective of testing is thus rather to improve
the software quality by finding faults in it.
Testing is an important activity of software
development, whose cost is usually estimated to about 40\% of
the total cost of software development, exceeding the cost
of code writing.

A test campaign for a program requires several steps:
design and development of test sets, execution and results
examination (or {\em oracle}).
Considering the cost of testing,
it is interesting to automate some of these steps.

For Java programs, the JUnit framework \cite{JUnit,BG98} allows 
the developer to write an oracle for each test case,
and to automatically execute test sets.
JUnit in particular permits to automatically regression
test several test sets.

If a formal specification is available, it can be translated into
assertions which can be checked at runtime, and thus serve 
as a test oracle. For instance, the DAISTS system \cite{GMH81}
compiles algebraic axioms of an abstract data type into consistency checks;
Rosenblum's APP pre-processor allows the programmer to write assertions for 
C programs \cite{Ros92}; the Eiffel {\em design and contract} approach 
integrates assertions in the programming language \cite{Mey88,Mey92}.

It is also interesting to automate the {\em development} of tests.
We can distinguish between two groups of strategies to produce
test sets: random and systematic strategies. Systematic 
strategies, such as functional testing or structural testing, 
consist in decomposing the input domain of the program in
several subdomains, often called ``partitions''. 

Many systematic strategies propose to derive test cases from 
a formal specification \cite{Gau95}. For instance, the Dick 
and Faivre method \cite{DF93}, which consists in constructing
a finite state automaton from the formal
specification and in selecting test cases as
paths in this automaton, has been used as
a basis by other approaches, in particular
Casting \cite{ABL97} or BZTT \cite{LPU02}.
BZTT uses B or Z specifications to
generate test cases which consist in placing the
system in a boundary state and calling an operation
with a boundary value.

In contrast to systematic methods, random testing generally 
does not use the program nor the specification to produce test sets.
It may use an {\em operational profile} of the program, 
which describe how the program is expected to be used.
The utility of random testing is controversial in the testing community. 
It is usually presented as the poorest approach for selecting test 
data \cite{Mye79}.
However, random testing has a few advantages which make us think that 
it could be a good complement to systematic testing:
\begin{itemize}
\item
   random testing is cheap and rather easy to implement. In particular, it can
   produce large or very large test sets;
\item
   it can detect a substantial
   number of errors at a low cost \cite{DN84,HT90,Nta01}.
\end{itemize}
Moreover, if an operational profile of the program is available,
\begin{itemize}
\item
   random testing allows early detection of the failures that are 
   most likely to appear when using the program \cite{FHLS98};
\item
   it can be used to evaluate the program {\em reliability} \cite{HT90,Ham94}.
\end{itemize}
However, partition testing can be much more effective at finding failures,
especially when the strategy defines some very small subdomains with a 
high probability to cause failures \cite{WJ91}. 

Among the best practices of {\em extreme programming}, or XP
\cite{Bec99,Bec00}, are {\em continuous testing} and
{\em code refactoring}.
While they are writing code, developers should write
corresponding unit tests, using a testing framework such as JUnit. 
Unit testing is often presented as a support to
refactoring: it gives the developer confidence that
the changes have not introduced new errors.
However, code refactoring often requires to
change some of the corresponding unit tests as well.
If a class has to be tested intensively, the 
amount of code corresponding to the tests often exceeds
the amount of code of the class. Both practices 
can therefore be hard to conciliate.

This report proposes to use random generation of 
tests to facilitate countinuous testing and code 
refactoring in the context of extreme programming.
We presents Jartege, a tool for random generation
of unit tests for Java classes specified in JML, 
which aims at easily producing numerous test cases,
in order to detect a substantial number of 
errors at a low cost.

The rest of the report is organized as follows.
Section \ref{Principles} introduces the approach.
Section \ref{CaseStudy}
presents a case study which consists in modeling
bank accounts. This case study will serve to illustrate the
use of our testing tool.
Section \ref{Jartege} presents our tool Jartege
and how it can be used to test the bank account case study.
Section \ref{ControlingRandom}
introduces more advanced features of Jartege, which
allow one to parameterize its random aspect.
Section \ref{ApplyingJartege}
shows the errors which are detected by test cases generated by
Jartege in the case study.
Section \ref{RelatedWork}
presents and compares related approaches.
Section \ref{Discussion} discusses some points about 
random generation of tests and draw future work
we intend to undertake around Jartege.

\section{Approach}%
\label{Principles}

JML (Java Modeling Language) is a
specification language for Java inspired by Eiffel, VDM and Larch,
which was designed by Gary Leavens and his colleagues \cite{JML,LBR03,LPC03}.
Several teams are currently still working on JML design and tools
around JML \cite{BCC03}. 
JML allows one to specify various assertions in particular
invariants for classes as well as pre- and postconditions
for methods. 

The JML compiler (\code{jmlc}) \cite{CL02a} translates JML
specifications into assertions checked at runtime.
If an assertion is violated, then a specific exception
is raised. In the context of a given method call, the JML compiler
makes a useful difference between an {\em entry precondition}
which is a precondition of the given method,
and an {\em internal precondition},
which is a precondition of an operation being called,
at some level, by the given method.

The JML-JUnit tool \cite{CL02b} generates
JUnit test cases for a Java program specified in JML,
using the JML compiler to translate JML 
specifications into test oracles. Test cases are produced 
from a test fixture and parameter values supplied by the user.

Our approach is inspired by the JML-JUnit tool: we propose 
to generate random tests for Java programs specified 
in JML, using this specification as a test oracle, in the JML-JUnit 
way. Our aim is to produce a big number of tests at a low cost,
in order to facilitate unit testing. 

To implement these ideas, we started developing a prototype tool,
called {\em Jartege} for {\em Java Random Test Generator}.
Jartege is designed to generate {\em unit tests} for Java classes
specified with JML. By unit tests, we here mean
tests for some operations in a single class or a small
cluster of classes.
In our context, a {\em test case} is a Java method which consists
of {\em a sequence of operation (constructor or method) calls}.

As in the JML-JUnit tool, we use the JML specification to assist test 
generation in two ways:
\begin{enumerate}
\item
   It permits the rejection of test sequences which contain an operation
   call which violates the operation {\em entry precondition}.
   We consider that these test sequences are not interesting 
   because they detect errors which correspond to a fault 
   {\em in the test program}. 

   (Although such sequences could be used to detect cases when a 
   precondition is too strong, this goal would prevent us from 
   producing long sequences of calls. Thus, we choose to trust the 
   specification rather than the code. We can note that if an operation
   uses a method whose precondition is too strong, this will produce 
   an {\em internal precondition} error and the sequence will {\em not}
   be rejected.)

\item
   The specification is also used as a test oracle:
   the test detects an error when another assertion (e.g. an invariant,
   a postcondition or an {\em internal precondition}) is violated.
   Such an error corresponds to a fault in the Java program or in
   the JML specification.
\end{enumerate}

Our work has been influenced by the Lutess tool \cite{Par96,BOR99},
which aims at deriving test data for synchronous programs, with various
generation methods, in particular a purely random generation
and a generation guided by operational profiles.
The good results obtained by Lutess, which won the best tool
award of the first feature interaction detection contest
\cite{BZ99}, have encouraged us to consider random generation of tests
as a viable approach.

\section{Case Study}%
\label{CaseStudy}

In this section, we present a case study
to illustrate the use of Jartege. This case study defines
{\em bank accounts} and some operations on these accounts.

\subsection{Informal Specification of the Bank Accounts}

We describe here an informal specification of bank accounts:

\begin{enumerate}
\item
   An account contains a certain available amount of money
   (its {\em balance}), and is associated with a minimum amount
   that this account may contain (the {\em minimum balance}).
\item
   It is possible to credit or debit an account. A debit operation
   is only possible if there is enough money on the account.
\item
   One or several last credit or debit operations may be cancelled.
\item
   The minimum balance of an account may be changed.
\end{enumerate}

\subsection{Bank Account Modeling}

\begin{figure}[|htbp]
\begin{center}
   \input{diag01.pstex_t}
\end{center}
\caption{UML diagram of the case study}\label{diag01}
\end{figure}
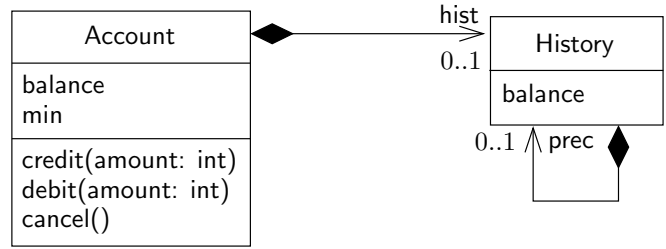

To represent accounts, we define a class \code{Account} with 
two attributes: \code{balance} and \code{min}
which respectively represent the balance and the minimum
balance of the account, and three methods: \code{credit}, \code{debit}
and \code{cancel}.
In order to implement the cancel operation, we associate
with each account an {\em history}, which is a linked list of
the previous balances of the account.
The class \code{History} has one attribute, \code{balance}, which
represents the balance of its associated account before the last
credit or debit operation. With each history is associated
its preceding history.
Figure~\ref{diag01} shows the UML class diagram of bank accounts.

\vspace*{2ex}

\subsection{JML Specification and Java Implementation}

We implement both class \code{Account} and \code{History} in Java,
and specify them with JML. We choose to write {\em lightweight
public specifications} to emphasize that these specifications are
destined for clients and need not be complete.

In order to forbid uncontrolled modifications
of the attributes, we declare them as private and define
associated access methods: \code{getBalance}, \code{getMin} and
\code{getHist} in class \code{Account} and \code{getBalance}
and \code{getPrec} in class \code{History}.
These methods are specified as {\em pure} methods in JML
because they are side effect free, which allows us to use them
in JML assertions.

The class \code{Account} has an invariant which specifies
that the balance of an account must always be greater
than the minimum balance.

\begin{program}
/* \Com{Class of bank accounts.} */
\mr{public} \mr{class} \id{Account} \lb
   /* \Com{Invariant of class Account.} */
   /*@ \jml{\mr{public} \mr{invariant} \id{getBalance(\,)} >= \id{getMin(\,)};} */

   \mr{private} \mc{int} \id{balance}; // \com{The balance of this account}
   \mr{private} \mc{int} \id{min};    // \com{The minimum balance}
   \mr{private} \id{History} \id{hist}; // \com{The history list of this account}

   /* \Com{The balance of this account.} */
   \mr{public} /*@\jml{\mr{ pure }}*/ \mc{int} \id{getBalance(\,)} \lb \mr{return} \id{balance}; \rb

   /* \Com{The history list of this account.} */
   \mr{public} /*@\jml{\mr{ pure }}*/ \id{History} \id{getHist(\,)} \lb \mr{return} \id{hist}; \rb

   /* \Com{The minimum balance of this account.} */
   \mr{public} /*@\jml{\mr{ pure }}*/ \mc{int} \id{getMin(\,)} \lb \mr{return} \id{min}; \rb
\end{program}
The constructor of class \code{Account} constructs an account
with the specified balance and the specified minimum balance.
Its precondition asserts that the specified balance is greater than
the specified minimum balance.

\begin{program}
   /* \Com{Constructs an account with the specified balance and} 
    * \Com{minimum balance.} */
   /*@ \jml{\mr{requires} \id{balance} >= \id{min};} */
   \mr{public} \id{Account\,(}\mc{int} \id{balance}, \mc{int} \id{min)} \lb
      \mc{this}.\id{balance} = \id{balance};
      \mc{this}.\id{min} = \id{min};
      \mc{this}.\id{hist} = \mc{null};
   \rb
\end{program}
As the minimum balance \code{min} is a private attribute, we have
to introduce a method to change its value. The method
\code{setMin\,(\mc{int} min)} sets the minimum balance to the
specified value. Its precondition asserts that the balance
is greater than the specified minimum value.
\begin{program}
   /* \Com{Sets the minimum balance to the specified value.} */
   /*@ \jml{\mr{requires} \id{getBalance\,(\,)} >= \id{min};} */
   \mr{public} \mc{void} \id{setMin\,(}\mc{int} \id{min)} \lb \mc{this}.\id{min} = \id{min}; \rb
\end{program}
The method \code{credit\,(\mc{int} amount)} credits the
account with the specified amount. Its precondition
requires the amount to be positive.
Its postcondition asserts that the new balance is the former balance
augmented by the specified amount, that a new history
is created with balance the former balance of the account and
with previous history the former history of the account.
Its exceptional postcondition asserts that the method should
never terminate abruptly.
\begin{program}
  /* \Com{Credits this account with the specified amount.} */
  /*@ \jml{\mr{requires} \id{amount} >= 0;}
   *@ \jml{\mr{ensures} \id{getBalance\,(\,)} == {\back}\id{old}\id{\,(getBalance\,(\,))} + \id{amount} &&}
   *@ \jml{       {\back}\mr{fresh\,}\id{(getHist\,(\,))} &&}
   *@ \jml{       \id{getHist\,(\,)}.\id{getBalance\,(\,)} == {\back}\mr{old}\id{\,(getBalance\,(\,))} &&}
   *@ \jml{       \id{getHist\,(\,)}.\id{getPrec\,(\,)} == {\back}\mr{old}\id{\,(getHist\,(\,))};}
   *@ \jml{\mr{signals} \id{(Exception} \id{e)} \mc{false};}
   */
  \mr{public} \mc{void} \id{credit}(\mc{int} \id{amount}) \lb
     \id{hist} = \mr{new} \id{History\,(balance}, \id{getHist\,(\,))};
     \id{balance} = \id{balance} + \id{amount};
  \rb
\end{program}
The debit operation, which is very similar to the credit operation, is 
not detailed here.
It has the additional precondition that the balance
decreased by the specified amount is greater than the minimum
balance. 

The method \code{cancel} cancels the last credit or debit
operation. Its precondition requires that the history is
not null, which means that at least one operation of credit
or debit has taken place since the account was created.
Its postcondition ensures that the balance and the history
of the account have been updated with their former values.
\begin{program}
  /* \Com{Cancels the last credit or debit operation.} */
  /*@ \jml{\mr{requires} \id{getHist\,(\,)} != \mr{null};}
   *@ \jml{\mr{ensures} \id{getHist\,(\,)} == {\back}\mr{old}\id{\,(getHist\,(\,)}.\id{getPrec\,(\,))} &&}
   *@ \jml{       \id{getBalance\,(\,)} == {\back}\mr{old}\id{\,(getHist\,(\,)}.\id{getBalance\,(\,))};}
   *@ \jml{\mr{signals} \id{(Exception} \id{e)} \mc{false};}
   */
  \mr{public} \mc{void} \id{cancel\,(\,)} \lb
     \id{balance} = \id{hist}.\id{getBalance\,(\,)};
     \id{hist} = \id{hist}.\id{getPrec\,(\,)};
  \rb
\rb // \com{End of class Account}
\end{program}
We do not define any JML assertion for the class \code{History}.
\begin{program}
/* \Com{Class of histories.} */
\mr{public} \mr{class} \id{History} \lb
  \mr{private} \mc{int} \id{balance};  // \com{The balance of this history.}
  \mr{private} \id{History} \id{prec}; // \com{The preceding history.}

  /* \Com{Constructs a history with the specified balance and preceding history.} */
  \mr{public} \id{History\,(}\mr{int} \id{balance}, \id{History} \id{prec)} \lb
     \mc{this}.\id{balance} = \id{balance}; \mc{this}.\id{prec} = \id{prec};
  \rb

  /* \Com{The balance of this history.} */
  \mr{public} /*@ \jml{\mr{pure}} */ \mc{int} \id{getBalance\,(\,)} \lb \mr{return} \id{balance}; \rb
  /* \Com{The preceding history.} */
  \mr{public} /*@ \jml{\mr{pure}} */ \id{History} \id{getPrec\,(\,)} \lb \mr{return} \id{prec}; \rb
\rb // \com{End of class History}
\end{program}

\section{Jartege}%
\label{Jartege}

Jartege (Java Random Test Generator) is a framework
for automatic random generation of unit tests for
Java classes specified with JML.
This approach consists in producing test programs
which are composed of test cases, each test case
consisting of randomly chosen sequences of
method calls for each class under test.
Each generated test program can be executed to test
the classes, and re-executed later on either after having
corrected some faults or for regression test.

The tool is designed to produce {\em unit tests}, i.e.
tests composed of calls of some methods which belong to a
few classes. As noticed in \cite{LTW00}, because
of complex dependences that exist between classes in
object-oriented programs, it is usually
not possible to test a method or a class in complete isolation.
Jartege thus is able to generate test cases which allow
the integration of several classes.

\subsection{Practical Use of Jartege}

Suppose we wish to generate tests for the classes \code{Account}
and \code{History}. We write the following Java program:

\begin{program}
\mr{import} \id{jartege}.*;
/** \Com{Jartege test cases generator for classes Account and History.} */
\mr{class} \id{TestGen} \lb
  \mr{public} \mr{static} \mc{void} \id{main\,(}\id{String[\,]} \id{args)} \lb
    // \com{Creates a class tester}
    \id{ClassTester} \id{t} = \mr{new} \id{ClassTester\,(\,)};
    // \com{Adds the specified classes to the set of classes under test}
    \id{t}.\id{addClass\,(}"\id{Account}"\id{)};
    \id{t}.\id{addClass\,(}"\id{History}"\id{)};
    // \com{Generates a test class TestBank, made of 100 test cases.}
    // \com{For each test case, the tool tries to generate 50 method calls.}
    \id{t}.\id{generate\,(}"\id{TestBank}", 100, 50\id{)};
  \rb
\rb
\end{program}

The main class of the Jartege framework is \code{ClassTester}.
This class must be instantiated to allow the creation of
test programs.

The method \code{addClass\,(String className)} adds the class
\code{className} to the set of classes under test. In this
example, we wish to generate tests for the classes
\code{Account} and \code{History}.

The method \code{generate\,(String className, \mc{int} numberOfTests,
\mc{int} numberOfMethodCalls)} generates a file \code{\mr{className}.java}
which contains a class called \code{\mr{className}}. This class
is composed of \code{\mr{numberOfTests}} test cases.
For each test case, the tool makes \code{\mr{numberOfMethodCalls}}
attempts to generate a method call of one of the classes under test.
Using the accessible constructors, the tool
constructs objects which serve as parameters for these method calls.

When the program \code{TestGen} is executed, it produces
a file \code{TestGen.java} which contains a main program.
This main program calls successively 100 test methods
\code{test1\,(\,)}, \code{test2\,(\,)} ... \code{test100\,(\,)}.
Each test method contains about 50 method calls.

While the program is generated, Jartege executes {\em on
the fly} each operation call, which allows it to eliminate calls
which violate the operation precondition. When this precondition
is strong, it may happen that the tool does not succeed in
generating a call for a given method, which explains that
{\em about} 50 method calls are generated.

\subsection{Test Programs Produced by Jartege}

A test program produced by Jartege is a class
with a \code{main} method which consists in
calling sequentially all generated test cases.
Each test case consists of a sequence of constructor and method
calls of the classes under test.
Here is an example of such a test case:

\begin{program}
// \com{Test case number 1}
\mr{public} \mr{void} \id{test1\,(\,)} \mr{throws} \id{Exception} \lb
  \mr{try} \lb
    \id{Account} \id{ob1} = \mr{new} \id{Account\,(1023296578, 223978640)};
    \id{ob1}.\id{debit\,(152022897)};
    \id{History} \id{ob2} = \mr{new} \id{History\,(1661966075, (History)} \mr{null)};
    \id{History} \id{ob3} = \mr{new} \id{History\,(-350589348, ob2)};
    \id{History} \id{ob4} = \id{ob2}.\id{getPrec\,(\,)};
    \mc{int} \id{ob5} = \id{ob3}.\id{getBalance\,(\,)};
    \id{ob1}.\id{cancel\,(\,)};
    // \com{...}
  \rb \mr{catch} \id{(Throwable} \id{\_except)} \lb
    \id{error\,(}\id{\_except}, 1\id{)};
  \rb
\rb
\end{program}

For each test method, if a JML exception is raised, then
an error message (coming from \code{jmlc}) is printed.
The test program terminates by printing an assessment of the test.
As an example, here is an excerpt of what is printed
with a generated program \code{TestBank.java}:

\begin{program}\sf
1) Error detected in class TestBank by method \persp{test2}:
    org.jmlspecs.jmlrac.runtime.\persp{JMLInvariantError}:
    By method "\persp{credit}@\persp{post}{\hlt}Account.java:79:18\hgt" of class "\persp{Account}" 
    for assertions specified at Account.java:\persp{11}:32 [...]
    at TestBank.test2(TestBank.java:138)
[...]
Number of tests: 100
Number of errors: 71
Number of inconclusive tests: 0
\end{program}

The program has detected 71 errors. The first error detected comes from 
a violation of the invariant
of class \code{Account} (specified line 11),
which happened after a \code{credit} operation.

The test program also indicates the number of {\em inconclusive}
tests. A test case is inconclusive when it does not allow one to
conclude whether the program behaviour is correct or not.
A test program generated by Jartege indicates that a test case
is inconclusive when it contains an operation call whose
{\em entry precondition} is violated.
As Jartege is designed to eliminate such operation calls, this situation
may only arise when the code or the specification of one of the classes
under test has been modified after the test file was generated.
A high number of inconclusive tests indicates that the test file
is no longer relevant.

\section{Controlling Random Generation}%
\label{ControlingRandom}

If we leave everything to chance, Jartege might
not produce interesting sequences of calls.
Jartege thus provides a few possibilities to parameterize
its random aspect. 
These features can be useful for stress testing, for instance
if we want to test more intensively a given method.
More generally, they allow us to define an {\em operational profile}
for the classes under test, 
which describe how these classes are likely to be used 
by other components.

\subsection{Weights}

With each class and operation of a class is associated
a weight, which defines the probability that a class will
be chosen, and that an operation of this class
will be called. In particular, it is possible to forbid to
call some operation by associating a null weight with it.
By default, all weights are equal to 1. A weight can be modified
by a weight change method. In particular:

\begin{itemize}
\item
   \code{changeAllMethodsWeight\,(String className, double weight)}
   changes the weight of all methods in the class \code{className}
   to the specified weight.

\item
   \code{changeMethodWeight\,(String className, String methodName, double weight)}
   changes the weight of the specified method(s) in the class \code{className}
   to the specified weight.

\item
   \code{changeMethodWeight\,(String className, String methodName,
   String\,[\,] signature, double weight)}
   changes the weight of the method by its name and its signature to
   the specified weight.
\end{itemize}

\subsection{Creation of Objects}

Objects creation is commanded by {\em creation probability
functions}, which define the probability of creating a new object
according to the number of existing objects of the class against
that of reusing an already created object.
If this probability is low, Jartege is more likely to
reuse an already created object than to construct a new one.
This allows the user either to create a predefined number of instances
for a given class, or on the opposite, to create numerous
instances for a class.

In the example of bank accounts, it is not very interesting to
create many accounts. It is possible to test the class
Account more efficiently for example by creating a unique account and
by applying numerous method calls to it.

The function
\code{changeCreationProbability(String className, CreationProbability
creationProbabilityFunction)} changes the creation probability function
associated with the specified class to the the specified creation
probability function.

The interface \code{CreationProbability} contains a unique method
\begin{center}
   \code{double theFunction(int nbCreatedObjects)} 
\end{center}
which must satisfy
the condition
\begin{center}
\begin{tabular}{l}
    \code{theFunction}(0) = 1; \\
    \code{theFunction}($n$) $\in$ [0, 1], $\forall n \geq 1$.
\end{tabular}
\end{center}

The class \code{ThresholdProbability} allows one to define
{\em threshold probability functions} whose value is 1 under
some threshold $s$ and 0 above.
\begin{center}
\begin{tabular}{l}
    \code{theFunction}($n$) = 1, if $n < s$; \\
    \code{theFunction}($n$) = 0, otherwise.
\end{tabular}
\end{center}
A threshold probability function with threshold $s$ allows one
to define at most $s$ instances of a given class.
We can for instance forbid the creation of more than one 
instance of \id{Account} by adding the following statement in the test generator:
\begin{program}
    \id{t}.\id{changeCreationProbability\,(}"\id{Account}", \mr{new} \id{ThresholdProbability\,(1))};
\end{program}

\subsection{Parameter Generation of Primitive Types}

When a method has a strong precondition, the probability that
Jartege, without any further indication, will generate a call
to this method which does not violate this precondition is
low. For primitive types, Jartege provides the possibility
to define generators for some parameters of a given method.

For example, the precondition of the \code{debit} operation
requires the parameter \code{amount} to be in range
\begin{center}
   [0, \code{getBalance\,(\,)} $-$ \code{getMin\,(\,)}]. 
\end{center}
Let us suppose the range is small, in other words that
the balance is closed to the minimum balance.
If Jartege chooses an amount to
be debited entirely randomly, this amount
is not likely to satisfy the method precondition.

Jartege provides a way of generating parameter values
of primitive types for operations. For this, we define a class
\code{JRT\_Account} as follows.

\begin{program}
\mr{import} \id{jartege}.\id{RandomValue};
\mr{public} \mr{class} \id{JRT\_Account} \lb
  \mr{private} \id{Account} \id{theAccount}; // \com{The current account}

  /* \Com{Constructor.} */
  \mr{public} \id{JRT\_Account\,(}\id{Account} \id{theAccount)} \lb \mc{this}.\id{theAccount} = \id{theAccount}; \rb

  /** \Com{Generator for the first parameter of operation debit\,(int).} */
  \mr{public} \mc{int} \id{JRT\_debit\_int\_1\,(\,)} \lb
     \mr{return} \id{RandomValue}.\id{intValue\,(0},
        \id{theAccount}.\id{getBalance\,(\,)} - \id{theAccount}.\id{getMin\,(\,))};
  \rb
\rb
\end{program}

The class \code{JRT\_Account} must contain a private field of type
\code{Account} which will contain the current object on which
an operation of class \code{Account} is applied.
A constructor allows Jartege to initialize this private field.
The class also contains one parameter generation method for
each parameter for which we specify the generation of values.
In the example, to specify the generation of the
first parameter of operation \code{debit\,(\mc{int} amount)}, 
we define the method \code{\mc{int} \id{JRT\_debit\_int\_1\,(\,)}}.
We use the signature of the operation in the name of the
method to allow overloading.
The method \code{RandomValue.intValue\,(\mc{int} min, \mc{int} max)}
chooses a random integer in range [\code{min}, \code{max}].

\subsection{Fixtures}

If we want to generate several test cases
which operate on a particular set of objects, we can write a
{\em test fixture}, in a similar way to JUnit.
A test fixture is a class which contains:
\begin{itemize}
\item
   attributes corresponding to the objects a test operates on;
\item
   an optional \code{setUp} method which defines the preamble of a test case
   (which typically constructs these objects);
\item
   an optional \code{tearDown} method which defines the postamble of a
   test case.
\end{itemize}

\section{Applying Jartege to the Case Study}%
\label{ApplyingJartege}

The 100 test cases generated by Jartege, showing 71 failures, 
revealed three different errors:
one error caused by a credit operation, and two 
errors caused by a cancel operation.
We extracted the shorter sequence of calls which resulted in each
failure and obtained the following results. We also
changed the parameter values and added some comments
for more readability.

\paragraph{Error 1.}
The credit operation can produce a balance inferior to the
previous balance because of an integer overflow.

\begin{program}
  \mr{public} \mc{void} \id{test1\,(\,)} \lb
    \id{Account} \id{ob1} = \mr{new} \id{Account\,(250000000, 0)};
    \id{ob1}.\id{credit\,(2000000000)}; // \com{Produces a negative balance,} 
  \rb                       // \com{below the minimum balance.}
\end{program}

\paragraph{Error 2.}
The cancel operation can produce an incorrect result if
it is preceded by a \code{setMin} operation which
changes the minimum balance of the account to a value
which is superior to the balance before cancellation.

\begin{program}
  \mr{public} \mc{void} \id{test11\,(\,)} \lb
    \id{Account} \id{ob1} = \mr{new} \id{Account\,(-50, -100)};
    \id{ob1}.\id{credit\,(100)};
    \id{ob1}.\id{setMin\,(0)};
    \id{ob1}.\id{cancel\,(\,)}; // \com{Restores the balance to a value}
  \rb              // \com{inferior to the minimum balance.}
\end{program}

\paragraph{Error 3.}
The third error detected is a combination of an overflow
on a debit operation
(similar to Error 1, which comes from an overflow on a credit operation)
and of the second error.

\begin{program}
  \mr{public} \mc{void} \id{test13\,(\,)} \lb
    \id{Account}\,\,\id{ob1}\,\,=\,\,\mr{new}\,\,\id{Account\,(-1500000000,\,-2000000000)};
    \id{ob1}.\id{debit\,(800000000)}; // \com{Produces a positive balance.}
    \id{ob1}.\id{setMin\,(0)};
    \id{ob1}.\id{cancel\,(\,)};  // \com{Restores the balance to a value}
  \rb               // \com{inferior to the minimum balance.}
\end{program}

We have the feeling that the three errors detected with test
cases generated by Jartege are not totally obvious, and could have
easily been forgotten in a manually developed test suite.
Errors 2 and 3 in particular require three method calls to be executed
in a specific order and with particular parameter values.

It must be noted that the case study was originally written
to show the use of JML to undergraduate students, without us
being aware of the faults.

\section{Comparison with Related Work}%
\label{RelatedWork}

Our work has been widely inspired by the JML-JUnit approach
\cite{CL02b}. The JML-JUnit tool generates test cases for
a method which consist of a combination of calls of this
method with various parameter values. The tester must
supply the object invoking the method and the parameter values.
With this approach, interesting values could easily be
forgotten by the tester. Moreover, as a test case only
consists of one method call, it is not possible to detect
errors which result of several calls of different methods.
At last, the JML-JUnit approach compels the user to
construct the test data, which may
require the call of several constructors.
Our approach thus has the advantage of being more
automatic, and of being able to detect more potential errors.


Korat \cite{BKM02} is a tool also based on the JML-JUnit
approach, which allows exhaustive testing of a method
for all objects of a bounded size. The tools automatically
construct all non isomorphic test cases and execute the
method on each test case.
Korat therefore has the advantage over JML-JUnit of being
able to construct the objects which invoke the method under test.
However, test cases constructed by Korat only consist
of one object construction and one method invocation on this
object.


Tobias \cite{Led02,MLB02} is a combinatorial testing tool for
automatic generation of test cases derived from a ``test pattern'',
which abstractly describes a test case.
Tobias was first designed to produce test objectives for
the TGV tool \cite{JM99} and was then adapted to produce
test cases for Java programs specified in JML and
for programs specified in VDM.
The main problem of Tobias is the combinatorial explosion
which happens if one tries to generate test cases which consist
of more than a couple of method calls.
Jartege was designed to allow the generation of long test
sequences without facing the problem of combinatorial explosion.

\section{Discussion and Future Work}%
\label{Discussion}

Jartege is only in its infancy and a lot of work remains to
be done.

Primitive values generation for methods parameters is
currently done manually by writing primitive parameters
generating methods.
Code for these methods could be automatically constructed
from the JML precondition of the method. This could consist
in extracting range constraints from the method precondition
and automatically produce a method which could generate
meaningful values for the primitive parameters.

Jartege easily constructs test cases which consist of
hundreds of constructors and methods calls.
It would be useful to develop a tool for extracting 
a minimum sequence of calls which results in a given failure.

We developed Jartege in Java, and we specified some
of its classes with JML. We applied Jartege to these
classes to produce test cases, which allowed us to
experiment our tool on a larger case study and
to detect a few errors. We found much easier to
produce tests with Jartege than to write unit tests
with JUnit or JML-JUnit. 
We intend to continue our work of specifying
Jartege in JML and testing its classes with itself.
We hope that this real case study will help us to
evaluate the effectiveness and scalability
of the approach.

A comparison of our work with other testing strategies 
still remains to be done. We can expect systematic methods, 
using for instance boundary testing such as BZTT \cite{LPU02},
to be able to produce more interesting test cases that ours.
Our goal is certainly not to pretend that tests produced
randomly can replace tests produced by more sophisticated
methods, nor a carefully designed test set written by an
experienced tester. 

Our first goal in developing Jartege was to help the 
developer to write unit tests for unstable Java classes, 
thus for ``debug unit testing''. 
It would also be interesting to use Jartege to evaluate 
the reliability of a stable component before it is released. 
Jartege provides some features to define
an operational profile of a component, which 
should allow statistical testing.
However, the definition of a correct operational profile,
especially in the context of object-oriented programming,
is a difficult task. Moreover, the relation between test sets
generated by Jartege and the reliability of a component requires 
more theoretical work, one difficult point being to take into account 
the state of the component. 

\section{Conclusion}%
\label{Conclusion}
This report presents Jartege, a tool for random generation
of unit tests for Java classes specified in JML. The aim of
the tool is to easily produce numerous test cases, in order to detect
a substantial number of errors without too much effort.
It is designed to produce automated tests, which can in part
replace tests written by the developer using for instance JUnit.
We think that the automatic generation of such unit tests
should facilitate continuous testing as well as code refactoring
in the context of extreme programming.

The JML specifications are used on the one hand to
eliminate irrelevant test cases, and on the other
hand as a test oracle. We think that the additional
cost of specification writing should be compensated
by the automatic oracle provided by the JML compiler,
as long as we wish to intensively test the classes.
Moreover, this approach has the advantage
of supporting the debugging of a specification along with the
corresponding program. This allows the developer to
increase his confidence in the specification and
to use this specification in other tools.

Most test generation methods are deterministic, while
our approach is statistical. We do not wish to oppose
both approaches, thinking that they both have
their advantages and drawbacks, and that a combination of both
could be fruitful.

At last, we found JML to be good language to start learning
formal methods. Its Java-based syntax makes it easy
to learn for Java programmers. As JML specifications
are included in Java source code as comments, it is
easy to develop and debug a Java program along with its
specification. Moreover, automatic test
oracles as well as automatic generation of test cases
are good reasons of using specification languages such as JML.

\bibliographystyle{alpha}
\bibliography{report}

\end{document}

%% file: diag01.pstex_t
\begin{picture}(0,0)%
\special{psfile=diag01.pstex}%
\end{picture}%
\setlength{\unitlength}{3947sp}%
\begingroup\makeatletter\ifx\SetFigFont\undefined
\def\x#1#2#3#4#5#6#7\relax{\def\x{#1#2#3#4#5#6}}%
\expandafter\x\fmtname xxxxxx\relax \def\y{splain}%
\ifx\x\y   
\gdef\SetFigFont#1#2#3{%
  \ifnum #1<17\tiny\else \ifnum #1<20\small\else
  \ifnum #1<24\normalsize\else \ifnum #1<29\large\else
  \ifnum #1<34\Large\else \ifnum #1<41\LARGE\else
     \huge\fi\fi\fi\fi\fi\fi
  \csname #3\endcsname}%
\else
\gdef\SetFigFont#1#2#3{\begingroup
  \count@#1\relax \ifnum 25<\count@\count@25\fi
  \def\x{\endgroup\@setsize\SetFigFont{#2pt}}%
  \expandafter\x
    \csname \romannumeral\the\count@ pt\expandafter\endcsname
    \csname @\romannumeral\the\count@ pt\endcsname
  \csname #3\endcsname}%
\fi
\fi\endgroup
\begin{picture}(4116,1540)(1639,-1840)
\put(1720,-864){\makebox(0,0)[lb]{\smash{\SetFigFont{10}{12.0}{rm}\id{balance}}}}
\put(2396,-554){\makebox(0,0)[b]{\smash{\SetFigFont{10}{12.0}{rm}\id{Account}}}}
\put(4822,-1222){\makebox(0,0)[rb]{\smash{\SetFigFont{10}{12.0}{rm}0..1}}}
\put(4588,-425){\makebox(0,0)[rb]{\smash{\SetFigFont{10}{12.0}{rm}\id{hist}}}}
\put(4605,-709){\makebox(0,0)[rb]{\smash{\SetFigFont{10}{12.0}{rm}0..1}}}
\put(5025,-1208){\makebox(0,0)[lb]{\smash{\SetFigFont{10}{12.0}{rm}\id{prec}}}}
\put(4733,-920){\makebox(0,0)[lb]{\smash{\SetFigFont{10}{12.0}{rm}\id{balance}}}}
\put(5190,-604){\makebox(0,0)[b]{\smash{\SetFigFont{10}{12.0}{rm}\id{History}}}}
\put(1720,-1706){\makebox(0,0)[lb]{\smash{\SetFigFont{10}{12.0}{rm}\id{cancel()}}}}
\put(1720,-1520){\makebox(0,0)[lb]{\smash{\SetFigFont{10}{12.0}{rm}\id{debit(amount: int)}}}}
\put(1720,-1333){\makebox(0,0)[lb]{\smash{\SetFigFont{10}{12.0}{rm}\id{credit(amount: int)}}}}
\put(1720,-1052){\makebox(0,0)[lb]{\smash{\SetFigFont{10}{12.0}{rm}\id{min}}}}
\end{picture}